\documentclass[twocolumn]{aastex62} 

\usepackage{soul} 
\usepackage{bm} 
\usepackage{amsmath} 
\PassOptionsToPackage{hyphens}{url}\usepackage{hyperref}

\graphicspath{{./}{Figures/}}

\shorttitle{PPTA~DR3: 3C~66B}
\shortauthors{Cardinal Tremblay et al.}

\begin{document}

\title{A Multimessenger Search for the Supermassive Black Hole Binary in 3C~66B with the Parkes Pulsar Timing Array}

\author[0000-0001-9852-6825]{Jacob Cardinal Tremblay}
    \email{jacob.cardinaltremblay@aei.mpg.de}
\affiliation{Max Planck Institute for Gravitational Physics (Albert Einstein Institute), 30167 Hannover, Germany}
\affiliation{Leibniz Universität Hannover, 30167 Hannover, Germany}

\author[0000-0003-3189-5807]{Boris Goncharov}
    \email{boris.goncharov@me.com}
\affiliation{Max Planck Institute for Gravitational Physics (Albert Einstein Institute), 30167 Hannover, Germany}
\affiliation{Leibniz Universität Hannover, 30167 Hannover, Germany}

\author[0000-0002-6428-2620]{Rutger van Haasteren}
\affiliation{Max Planck Institute for Gravitational Physics (Albert Einstein Institute), 30167 Hannover, Germany}
\affiliation{Leibniz Universität Hannover, 30167 Hannover, Germany}

\author[0000-0002-8383-5059]{N.~D.~Ramesh Bhat}
\affiliation{International Centre for Radio Astronomy Research, Curtin University, Bentley, WA 6102, Australia}

\author[0000-0001-7016-9934]{Zu-Cheng Chen}
\affiliation{Department of Physics and Synergetic Innovation Center for Quantum Effects and Applications, Hunan Normal University, Changsha, Hunan 410081, China}
\affiliation{Institute of Interdisciplinary Studies, Hunan Normal University, Changsha, Hunan 410081, China}

\author[0000-0003-3432-0494]{Valentina Di~Marco}
\affiliation{School of Physics and Astronomy, Monash University, Clayton, VIC 3800, Australia}
\affiliation{ARC Centre for Excellence for Gravitational Wave Discovery (OzGrav)} 
\affiliation{Australia Telescope National Facility, CSIRO, Space and Astronomy, PO Box 76, Epping, NSW 1710, Australia}

\author[0000-0002-7191-2301]{Satoru Iguchi}
\affiliation{National Astronomical Observatory of Japan, 2-2-1 Osawa, Mitaka, Tokyo 181-8588, Japan}

\author[0009-0001-5071-0962]{Agastya Kapur}
\affiliation{Department of Mathematics and Physical Sciences, Macquarie University, NSW 2109, Australia}    \affiliation{Australia Telescope National Facility, CSIRO, Space and Astronomy, PO Box 76, Epping, NSW 1710, Australia}

\author[0009-0009-9142-6608]{Wenhua Ling}
\affiliation{Australia Telescope National Facility, CSIRO, Space and Astronomy, PO Box 76, Epping, NSW 1710, Australia}

\author[0000-0001-5131-522X]{Rami Mandow}
\affiliation{Department of Mathematics and Physical Sciences, Macquarie University, NSW 2109, Australia}
\affiliation{Australia Telescope National Facility, CSIRO, Space and Astronomy, PO Box 76, Epping, NSW 1710, Australia}

\author[0009-0001-5633-3512]{Saurav Mishra}
\affiliation{Centre for Astrophysics and Supercomputing, Swinburne University of Technology, Hawthorn, VIC, 3122, Australia}
\affiliation{ARC Centre for Excellence for Gravitational Wave Discovery (OzGrav)} 
\affiliation{Australia Telescope National Facility, CSIRO, Space and Astronomy, PO Box 76, Epping, NSW 1710, Australia}

\author[0000-0002-2035-4688]{Daniel J. Reardon}
\affiliation{Centre for Astrophysics and Supercomputing, Swinburne University of Technology, Hawthorn, VIC, 3122, Australia}
\affiliation{ARC Centre for Excellence for Gravitational Wave Discovery (OzGrav)} 

\author[0000-0002-7285-6348]{Ryan M. Shannon}
\affiliation{Centre for Astrophysics and Supercomputing, Swinburne University of Technology, Hawthorn, VIC, 3122, Australia}
\affiliation{ARC Centre for Excellence for Gravitational Wave Discovery (OzGrav)} 

\author{Hiroshi Sudou}
\affiliation{National Institute of Technology, Sendai College, 48 Nodayama, Medeshima-Shiote, Natori, Miyagi 981-1239, Japan}

\author[0000-0001-9782-1603]{Jingbo Wang}
\affiliation{Institute of Optoelectronic Technology, Lishui University, Lishui 323000, China}

\author[0009-0001-8885-5059]{Shi-Yi Zhao}
\affiliation{School of Physics and Astronomy, Beijing Normal University, Beijing 100875, China}
\affiliation{Department of Physics, Faculty of Arts and Sciences, Beijing Normal University, Zhuhai 519087, China}

\author[0000-0001-7049-6468]{Xing-Jiang Zhu}
\affiliation{Department of Physics, Faculty of Arts and Sciences, Beijing Normal University, Zhuhai 519087, China}
\affiliation{Institute for Frontier in Astronomy and Astrophysics, Beijing Normal University, Beijing 102206, China}

\author[0000-0002-9583-2947]{Andrew Zic}
\affiliation{Australia Telescope National Facility, CSIRO, Space and Astronomy, PO Box 76, Epping, NSW 1710, Australia}
\affiliation{ARC Centre for Excellence for Gravitational Wave Discovery (OzGrav)}

\begin{abstract}
 
A subparsec supermassive black hole binary (SMBHB) at the center of the galaxy 3C~66B is a promising candidate for continuous gravitational-wave searches with pulsar timing arrays (PTAs). 
In this work, we search for such a signal in the third data release of the Parkes Pulsar Timing Array. 
Matching our priors to estimates of binary parameters from electromagnetic observations, we find a log Bayes factor $\ln \mathcal{B}= - 0.0027(7)$, highlighting that the source can be neither confirmed nor ruled out. 
We place upper limits at $95\%$ credibility on the chirp mass $\mathcal{M} < 6.90 \times 10^{8}\ M_{\odot}$, and on the characteristic strain amplitude  $\textrm{log}_{10}(h_0)< -14.44$. 
This partially rules out the parameter space suggested by electromagnetic (EM) observations of 3C~66B. 
We also independently reproduce the calculation of the chirp mass with the 3 mm flux monitor data from the unresolved core of 3C~66B.
Based on this, we outline a new methodology for constructing a joint likelihood of EM and gravitational-wave data from SMBHBs.
Finally, we suggest that targeted searches may allow firmly established SMBHB candidates to be treated as standard sirens, for complementary constraints on the Universe expansion rate. 
\end{abstract}



\section{\label{sec:intro} Introduction}

Pulsar timing arrays (PTAs) are experiments for long-term monitoring of millisecond pulsar pulse arrival times with the primary goal of detecting nanohertz-frequency gravitational waves (GWs).  
Recent PTA data show evidence for a nanohertz-frequency gravitational wave background~\citep[GWB;][]{GoncharovShannon2021, ng12_gwb, ChenCaballero2021, ipta_dr2_gwb, ng15_gwb, epta_dr2_gwb, ReardonZic2023b}. 
The background origin is consistent with the superposition of adiabatically inspiraling supermassive black hole binaries ~\citep[SMBHBs;][]{GoncharovSardana2025a}.
Provided confident detection of the background, the detection of a continuous gravitational wave (CGW) from an individual SMBHB is the next milestone for PTAs. 
Such an observation would decisively resolve the final parsec problem, which represents difficulties in modeling the evolution of SMBHBs to subparsec separations \citep{MilosavljevicMerritt2003}. 
When the host galaxy of CGW source is known, the detection unlocks multimessenger astronomy and cosmology with PTAs \citep{Schutz1986, HolzHughes2005, CharisiTaylor2022}. 

One of the most promising sources of CGWs is the SMBHB candidate in the center of galaxy 3C~66B. 
By using long baseline interferometry, \citet{SudouIguchi2003} found evidence for elliptical binary orbital motion in this radio galaxy's core and inferred a GW frequency of $f_{\textrm{GW}} = 60.4 \pm 1.73$ nHz. 
Further observations by \citet{IguchiOkuda2010} found that the binary has a chirp mass $\mathcal{M}=7.9^{+3.8}_{-4.5} \times 10^{8}\ M_{\odot}$, primarily through measuring the orbital radius of the electromagnetic (EM)-emitting black hole and the semi-major axis of an SMBHB.
Galaxy 3C~66B is located at a redshift $z = 0.02126$ \citep{vandenBoschGebhardt2015}, which corresponds to a luminosity distance $D_{\text{L}} = 94.17^{+0.20}_{-0.20}$~Mpc~\citep{PlanckCollaborationAghanim2020}. 

Based on the above, the expected strain amplitude $h_0$ of a CGW from 3C~66B is $7.3^{+6.8}_{-5.8} \times 10^{-15}$. 
This is the highest known strain amplitude of CGW sources in the PTA frequency band, and is within reach for PTA experiments. 
A summary of the expected properties of the SMBHB in 3C~66B is shown in Table \ref{tab:3c66b}. 
Because the properties in the table are inferred from EM observations, we will also refer to this model of the SMBHB in 3C~66B as the ``EM model". 

\begin{table}[!htb]
\caption{Summary of the Properties of the Supermassive Black Hole Binary Candidate in 3C~66B.}
\label{tab:3c66b}
\centering

\small
\setlength{\tabcolsep}{3pt}      
\renewcommand{\arraystretch}{1.2}
\sloppy                         

\begin{tabular}{lll}
\hline
 & Value & \parbox[t]{0.32\linewidth}{Reference} \\ \hline \hline
$\mathcal{M}$ ($M_{\odot}$) & $7.9^{+3.8}_{-4.5} \times 10^{8}$ & \citet{IguchiOkuda2010} \\
$f_{\textrm{GW}}$ (nHz) & $60.4 \pm 1.73$ & \citet{SudouIguchi2003} \\
$D_\text{L}$ (Mpc) & $94.17 \pm 0.20$ & \citet{vandenBoschGebhardt2015} \\
\shortstack[l]{R.A. (HH:MM:\\SS.SSS)} & 02:23:11.4110 & \citet{HuntJohnson2021} \\
\shortstack[l]{Decl. (DD:MM:\\SS.SSS)} & +42:59:31.385 & \citet{HuntJohnson2021} \\
$h_0$ & $7.3^{+6.8}_{-5.8} \times 10^{-15}$ & \citet{IguchiOkuda2010} \\
\hline
\end{tabular}
\end{table}

It is worth noting that \citet{IguchiOkuda2010} revised the chirp mass of the SMBHB in 3C~66B inferred in the earlier work of \citet{SudouIguchi2003}. 
The initial model of the SMBHB at the center of 3C~66B proposed by \citet{SudouIguchi2003} suggested a very high chirp mass of $1.3 \times 10^{10}\ M_{\odot}$. 
\citet{JenetLommen2004} have ruled out this model based on 7 yr of timing from PSR~B1855+09 \citep{KaspiTaylor1994}. 
Assuming a circular SMBHB in 3C~66B, they placed an upper limit of $7 \times 10^9\ M_{\odot}$ on its chirp mass. 
This upper limit does not exclude the model by \citet{IguchiOkuda2010}, which motivates further studies.

Following the revision of the chirp mass of the SMBHB candidate by \citet{IguchiOkuda2010}, a number of searches for a CGW from 3C~66B with PTAs have been performed. 
The authors of \citet{ng11_3c66b} performed a search for an SMBHB in a circular orbit with the 11 yr dataset of the North American Nanohertz Observatory for Gravitational Waves (NANOGrav). 
An upper limit on the chirp mass of $1.65 \times 10^9\ M_{\odot}$ was placed. 
A subsequent search was then performed for both circular and eccentric SMBHBs using the NANOGrav 12.5 yr dataset. 
The search for a circular binary~\citep{ng12_allsky_cw} constrained the mass to $1.41 \times 10^9\ M_{\odot}$. 
The search, which included eccentricity in the SMBHB system \citep{ng12_3c66b_eccentric}, placed their tightest constraints on the upper limit of the chirp mass with the value of $1.81 \times 10^9\ M_{\odot}$. The results from the 12.5 yr NANOGrav dataset are all still compatible with the \citet{IguchiOkuda2010} model, yet claim no detection, which motivates further searches for the source in PTA datasets. 
Similar to the results shown later in our work, the NANOGrav 15 yr dataset places a limit on the chirp mass at $95\%$ credibility as $9.54 \times 10^8~M_\odot$~\citep{ng15_cw_targeted}. On this same dataset, \citet{TianBi2025} found that for an eccentric model, the upper limit on the chirp mass is of $1.82 \times 10^{9}~M_{\odot}$.

In this work, we perform a search for a CGW from an SMBHB with a circular orbit in 3C~66B using the third data release (DR3) of the Parkes Pulsar Timing Array ~\citep[PPTA;][]{ZicReardon2023}. 
It is the first targeted search for 3C~66B with PPTA data. 
We expect the search based on these data to provide better constraints on a CGW from 3C~66B, thanks to an increased time span exceeding 18 yr in 13 pulsars \citep{MingarelliLazio2017}. 
As motivated by~\citet{LiuVigeland2021}, the targeting is performed by setting the sky position as a fixed parameter, as well as setting priors for other parameters based on the EM observations summarized in Table~\ref{tab:3c66b}. 
We employ both Bayesian and frequentist methods to calculate the significance of the hypothesis that an SMBHB with subparsec separation resides in 3C~66B. 
Finally, we discuss the current feasibility of cosmology with PTAs and produce constraints on the Hubble–Lemaître constant, assuming that the EM model is valid.

The rest of the paper is organized as follows. In Section~\ref{sec:dr3}, we introduce the dataset from DR3 of the PPTA. In Section~\ref{sec:method}, we introduce the signal model, the Bayesian analysis methodology, and our noise models. 
In Section~\ref{sec:results}, we present our results and discuss the model selection, upper limits on the chirp mass, upper limits on the characteristic strain amplitude, upper limits as a function of CGW frequency, and the frequentist analysis results.
In Section~\ref{sec:discussion}, we discuss new methods for a joint-likelihood analysis of 3C~66B and whether targeted searches in PTAs can be used for cosmology experiments.
In Section \ref{sec:conclusion}, we present our conclusions.

\section{Data}\label{sec:dr3}

The PPTA DR3 dataset~\citep{ZicReardon2023} is based on monitoring of $N_\text{psr}=32$ millisecond radio pulsars.
The total observing span of PPTA DR3 is $T_\text{obs}=18$~yr. 
The first observations took place on 2004 February 6, and the last observations took place on 2022 March 8. 
The observations were conducted using the 64 m Parkes `Murriyang' radio telescope in Australia, and observations were taken for each pulsar approximately every 3 weeks. 
The most recent 3 yr of observations in DR3 are carried out using the ultra-wide bandwidth, low-frequency receiver~\citep[UWL;][]{HobbsManchester2020}. In this work, only the data from 31 pulsars are used. PSR~J1741+1351 is excluded from the analysis due to the fact that it was only added to the PPTA observations after the commissioning of the UWL and was only observed with low priority.
With only 16 observations, this pulsar is not sensitive to any CGW signal and will therefore not contribute to the results.

As with any PTA data, PPTA~DR3 is a set of pulse arrival time observations, including observation details such as radio frequency and backend-receiver information, as well as the pulsar ephemeris, known as the timing model. 
A timing model predicts pulse arrival times based on the pulsar's specific spin, astrometry, properties of the interstellar plasma along the line of sight, and the binary parameters of the pulsar when applicable. 
We use the \texttt{DE440} ephemeris~\citep{ParkFolkner2021} to obtain pulse arrival times in an inertial frame centered at the solar system barycenter (SSB). 

Additionally, we follow~\citet{IguchiOkuda2010} by employing the same millimeter-wavelength flux variation data for 3C~66B. 
There are two components of this data. 
The first component of the data is collected at $93.7$~GHz using the Nobeyama Millimeter Array of the National Astronomical Observatory of Japan.
The second component of the data is collected at $86.2$~GHz using the Plateau de Bure Interferometer of the Institut de Radioastronomie Millim\'etrique.

\section{\label{sec:method}Methodology}

\subsection{\label{sec:method:signal}Signal Model }

We use the CGW signal model for nonspinning SMBHBs in circular\footnote{\citet{IguchiOkuda2010} point out that 3C~66B may be in a circular orbit.} orbits at zeroth post-Newtonian order described in \cite{ng11_allsky_cw} and \cite{ng11_3c66b,ng12_allsky_cw}. 
More details of the model are provided in Appendix~\ref{sec:appendix:cw}.
The CGW model for metric perturbations at SSB (the Earth term) is based on eight parameters:
\begin{equation}
     \{ \theta, \phi, f_\text{GW}, \Phi_0, \psi, \iota, \mathcal{M}, (D_{\text{L}} \wedge h_0) \}
\end{equation}
where $ ( \theta, \phi  )$ is the sky position, $f_\text{GW}$ is the GW frequency, $\Phi_0$ is the orbital phase, $\psi$ is the GW polarization angle, $\iota$ is the binary inclination angle, $\mathcal{M}$ is the chirp mass, $D_{\text{L}}$ is the luminosity distance and $h_0$ is the gravitational strain amplitude.
At zeroth post-Newtonian order, we compute the amplitude as 
\begin{equation}
\label{eq:strain_amp}
    h_0 = \frac{2 (G \mathcal{M})^{5/3}  ( \pi f_\text{GW}  )^{2/3}}{D_{\text{L}} c^4},
\end{equation}
unless it is provided as a free parameter (hence, $D_\text{L} \wedge h_0$). 
The full CGW model is based on metric perturbations at both the SSB and each pulsar (pulsar terms). 
This requires additional $2 N_\text{psr}$ parameters $\{L_i,\Phi_i\}$, $i=(1,N_\text{psr})$. 
They are the pulsar distance and the orbital phase at the pulsar, respectively. 
We provide more information on the pulsar distances in Appendix~\ref{sec:appendix:pdist}.
Searching only for the Earth term is sufficient for detection and upper limits, whereas including the pulsar term enhances the search sensitivity and remains important for parameter estimation when the signal is detected. 

\begin{figure}[h]
     \centering
     \includegraphics[width=\linewidth]{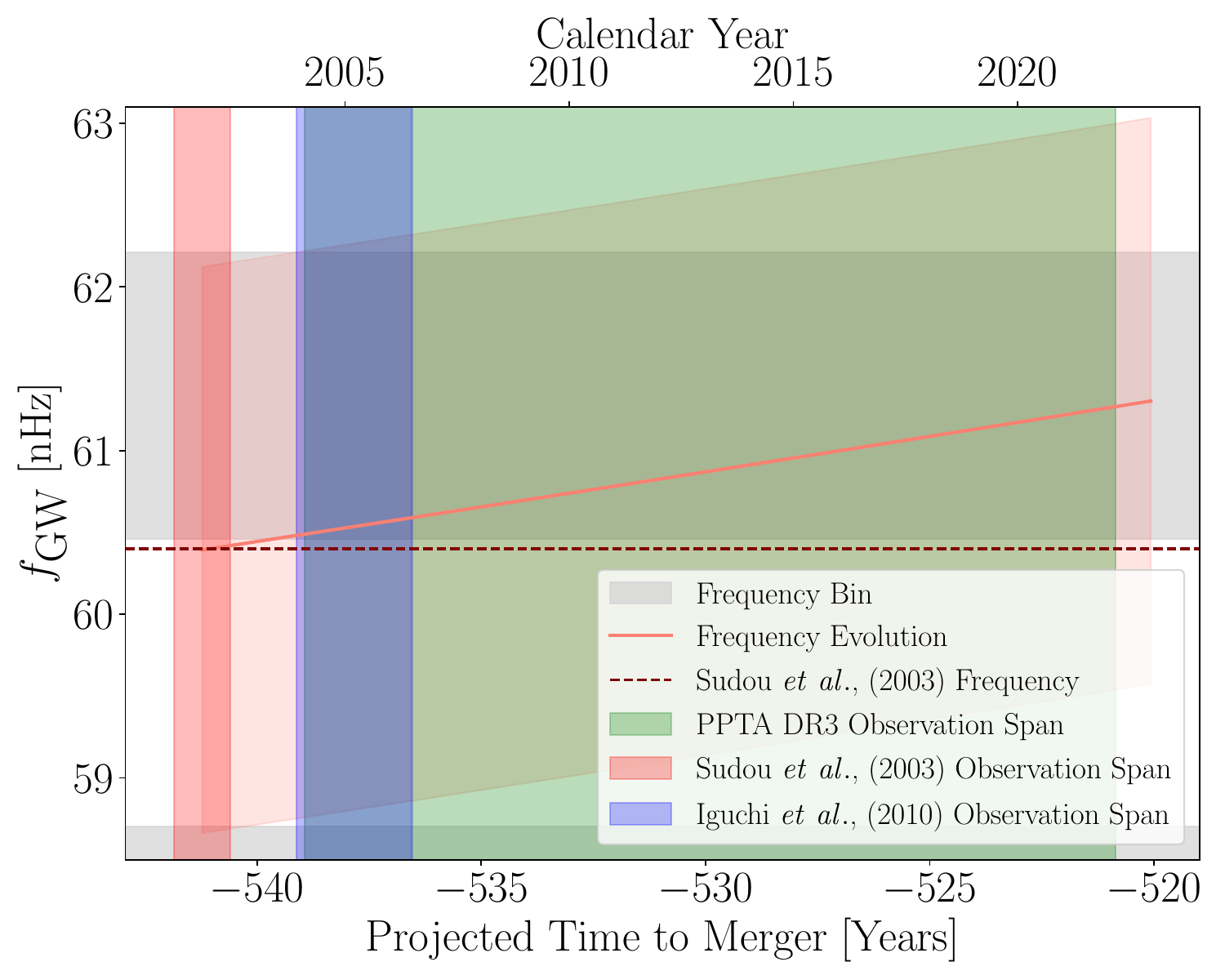}
     \caption{The evolution of the CGW frequency of 3C~66B at the SSB, and the associated uncertainty from~\citet{SudouIguchi2003} is shown by the pink diagonal line.  
     The dashed red line and the shaded red region show inferred frequency and their measurement uncertainty, respectively.
     The vertical red band represents the observing time frame of~\citet{SudouIguchi2003}. 
     The blue shaded region represents the observing time frame of~\citet{IguchiOkuda2010}. 
     The green shaded region represents the observing time frame of the PPTA~DR3 dataset. 
     The gray shaded region represents one frequency bin in PPTA~DR3 based on $T^{-1}_\text{obs}$.} 
     \label{fig:Freq_Evolution}
 \end{figure}

We model the evolution of CGW frequency on the timescales of the light travel time between the pulsars and Earth. 
However, for simplicity, we neglect frequency evolution on the timescale of PPTA~DR3. 
This approach is referred to as the stationary phase approximation. 
To demonstrate the validity of this approach, we employ an equation for the evolution of CGW frequency as a function of time at zeroth post-Newtonian order from~\citep{gw170817}:
\begin{equation}
    \label{eq:freqev}
    f_{\text{GW}}^{-8/3} ( t  ) = \frac{ ( 8\pi  )^{8/3}}{5}  \bigg(\frac{G\mathcal{M}}{c^3} \bigg)^{5/3} (t - t_\text{c}),
\end{equation}
where $f_{\text{GW}}$ is the GW frequency, $G$ is Newton's constant, $c$ is the speed of light, and $t_\text{c}$ is the time of coalescence. 
We present the expected frequency evolution of 3C~66B at SSB in Figure~\ref{fig:Freq_Evolution}.
Based on the EM model for 3C~66B, a circular SMBHB is only expected to evolve by about $0.04$ $\text{nHz}\ \text{yr}^{-1}$.
In the 20 yr since the initial frequency was determined by~\citet{SudouIguchi2003}, the frequency may have increased by $0.86$ nHz. 
This width does not exceed the frequency resolution of PPTA~DR3 of $T^{-1}_\text{obs} = 1.75$~nHz. 
Therefore, according to the model provided by \citet{SudouIguchi2003}, the stationary phase approximation is sufficient for our analysis. 

\subsection{\label{sec:method:bayesian}Bayesian Inference}

We perform parameter estimation and model selection by computing posteriors, $\mathcal{P}(\bm{\theta}|\bm{\delta t}) = \mathcal{Z}^{-1} \mathcal{L}(\bm{\delta t}|\bm{\theta}) \pi(\bm{\theta})$, where $\mathcal{L}(\bm{\delta t}|\bm{\theta})$ is the likelihood of data time series vector $\bm{\delta t}$ given a vector of model parameters $\theta$ for signals and noise, and $\pi(\bm{\theta})$ is the prior. 
The constant $\mathcal{Z}$ is the integral of the numerator over $\bm{\theta}$. 
It is also referred to as the evidence or the fully marginalized likelihood.
We neglect $\mathcal{Z}$ in our parameter estimation and obtain posterior samples using parallel-tempered Markov-Chain Monte-Carlo~\citep{EllisvanHaasteren2019}. 
The model selection follows from computing the Bayes factor, $\mathcal{B}=\mathcal{Z}/\mathcal{Z}_\varnothing$, where $\mathcal{Z}_\varnothing$ is the evidence of the noise model and $\mathcal{Z}$ is the evidence of the signal model, which includes both the noise and the CGW. 
Instead of computing $\mathcal{Z}$ separately for the two models, we compute the Bayes factor directly using the product-space sampling method~\citep{HeeHandley2016}.

We use the multivariate Gaussian likelihood~\citep{vanHaasterenLevin2009} 
\begin{equation}
\label{eq:likelihood}
    \mathcal{L}(\bm{\delta t} | \bm{\theta}) = \frac{\exp\bigg(  -\frac{1}{2} (\bm{\delta t} - \bm{\mu})^\text{T} \bm{C}^{-1}(\bm{\delta t}-\bm{\mu}) \bigg)}{\sqrt{\det\left( 2\pi \bm{C}\right)}},
\end{equation}
where vector $\bm{\mu}$, a function of $\bm{\theta}$, represents a prediction of deterministic contributions to pulse arrival times $\bm{\delta t}$, which includes a CGW model, as well as deterministic noise transients such as exponential dips in the pulse arrival times~\citep{GoncharovReardon2021}.
The form of our covariance matrix $\bm{C}$ and the methods for computing it are described in detail in~\citet{ng11_gwb}. 
In short, $\bm{C}=\bm{N} + \bm{T} \bm{B} \bm{T}^\text{T}$, where the block diagonal matrix $\bm{N}$ represents temporally uncorrelated ``white'' noise, $\bm{T}$ is the design matrix for reduced-rank modeling~\citep{LentatiAlexander2013}, and $\bm{B}$ is the prior covariance matrix of reduced-rank coefficients $\bm{b}$. 
The likelihood is marginalized over $\bm{b}$, which determines the time series realization $\bm{T} \bm{b}$. 
Thus, the design matrix $\bm{T}$ is a mapping from the parameter space $\bm{b}$ to the time series. 
It is made up of three blocks. 
Block $\bm{M}$ maps timing model parameters to the time series. 
Block $\bm{F}$ maps Fourier sine and cosine amplitudes of time-correlated ``red'' noise to the time series. 
Block $\bm{U}$ maps delays in the arrival times at different observing epochs due to pulse jitter to all arrival times. 
Coefficients $\bm{b}$ are assumed to be samples from the zero-mean Gaussian distribution described by the covariance matrix $\bm{B}$. 
For the coefficients of the pulsar timing model, we assume an uninformative prior with a standard deviation of $10^{40}$ in respective units, which is much greater than the typical coefficient value.

\subsection{\label{sec:method:noise}Noise Model}

Noise modeling in PTAs is crucial for correct astrophysical interpretability of the GW searches~\citep{GoncharovSardana2025b,HazbounSimon2020, LarsenMingarelli2024, ReardonZic2023b}. 
Broadly, modeled PTA noise is categorized into temporally uncorrelated (white) noise, temporally correlated (red) noise with the power spectral density modeled as a power law, as well as noise transients. 
We use the noise model from~\citet{ReardonZic2023b}. 
In addition to this, we model the contribution of the common-spectrum noise process associated with the GWB~\citep{GoncharovThrane2022}. 
It is shown in~\cite{ZhaoChen2025} that not modeling this process leads to false positives. 
\citet{ng15_cw} suggest that further modeling of the~\citet{HellingsDowns1983} correlations may be necessary for the data where they are evident. 
We do not model these correlations in this work; however, it may be required for future CGW searches with the PPTA data. 
Furthermore, we impose a minimum allowed value of the red noise power spectral density of $T_\text{obs} \times 10^{-20}~\text{s}^3$, which is well below the noise in PPTA~DR3. 
This is done to avoid numerical errors in the likelihood.

\subsection{\label{sec:method:3mm}Analysis of the 3 mm Data}
We perform a Bayesian reanalysis of the 3C 66B 3 mm flux measurements rather than using the original frequentist results from \citet{IguchiOkuda2010}. This allows a direct comparison between the GW posteriors and the EM posterior.
We employ the univariate Gaussian likelihood, $\mathcal{L}(\bm{S}|\bm{\theta}_\text{EM})$, where $\bm{S}$ is a vector of flux measurements in units of mJy.
The general form is the same as Equation~\ref{eq:likelihood}, except that the covariance matrix is diagonal with elements $\bm{\sigma}_\text{EM}^2 = \bm{\sigma}_S^2 e_\text{f}^2 + e_\text{q}^2$. 
Here, $\bm{\sigma}_S$ is the flux measurement uncertainty. 
White noise parameters $(e_\text{f},e_\text{q})$ are the error factor and the error added in quadrature, respectively. 
The analog of $\bm{\mu}$ from Equation~\ref{eq:likelihood} for the EM data is the sine cubic function from~\cite{IguchiOkuda2010}:
\begin{equation}
\label{eq:em_model}
\begin{split} 
    \bm{S}(t) = A_\text{EM} \sin(2 \pi f_\text{EM} t + \phi_\text{EM}) + \\ + a_\text{EM}(t-t_0)^3
    + b_\text{EM}(t-t_0)^2 + \\ + c_\text{EM}(t-t_0)  + d_\text{EM}
\end{split} 
\end{equation}
In total, $\bm{\theta}_\text{EM}$ is the set of parameters ($e_\text{f}$, $e_\text{q}$, $A_\text{EM}$~[mJy], $f_\text{EM}$~[Hz], $\phi_\text{EM}$, $t_0$~[s], $a_\text{EM}$~[mJy$\text{s}^{-3}$], $b_\text{EM}$~[mJy$\text{s}^{-2}$], $c_\text{EM}$~[mJy$\text{s}^{-1}$], $d_\text{EM}$~[mJy]).

The chirp mass is calculated from $\bm{\theta}_\text{EM}$ and Equation~\ref{eq:em_model} as follows. 
\citet{IguchiOkuda2010} suggest that the sine component is due to the Doppler boosting of the flux from the core of 3C~66B due to binary motion around the center of mass, whereas the cubic component may be responsible for the nonthermal radiation in the knots of the jet in the core. 
From the frequency of the sine component $f_\text{EM}$, the SMBHB orbital frequency $f_\text{k}$ from~\citet{SudouIguchi2003}, and the apparent boost in the relativistic jet $\beta_\text{app}$~\citep{IguchiOkuda2010}, we compute the jet outflow boost $\beta$ and the inclination angle $\iota$ between the jet axis and the line of sight.
The flux ratio from Doppler boosting \citep{RiegerMannheim2000, DePaolisIngrosso2002, DePaolisIngrosso2004} is $\bm{S}_\text{r}=\max(\bm{S})/\min(\bm{S})$, where $\max(\bm{S})=\bar{S}+A_\text{EM}$ and $\min(\bm{S})=\bar{S}-A_\text{EM}$ are computed from the mean flux $\bar{S}$.
From the above, assuming a power-law relation between the Doppler factor and $\bm{S}_\text{r}$ with spectral index $\alpha$, we calculate the SMBHB orbital radius. 
Considering that the jet originates from one of the black holes in a binary, but not two, and from the total mass of the binary $M_\text{tot}$, which can be computed based on the galaxy velocity dispersion relation, the chirp mass $\mathcal{M}$ is obtained. 
To sum up, $\mathcal{M} = \mathcal{G}(\theta_\text{EM},\bm{\varphi})$, where $\bm{\varphi}=(f_\text{k}$, $\beta_\text{app}$, $M_\text{tot}$, $\alpha)$ and $\mathcal{G}$ is a function with the aforementioned operations. 
For more details, please refer to Appendix~\ref{sec:appendix:em} and~\citet{IguchiOkuda2010}.

\section{\label{sec:results} Results}

\subsection{\label{sec:results:model_sel} Model Selection}

We first perform model selection by calculating the Bayesian evidence in favor of an SMBHB at 3C~66B against the null hypothesis. 
For this calculation, we match our prior on the CGW parameters to the results of EM searches. 
This approach allows for precise confirmation or exclusion of the EM model. 
In particular, for the $\log_{10}$ chirp mass, we use a skewed Normal prior with hyperparameters for the skewness, mean, and the standard deviation which have values of $-3.08$, $9.12~ [\log_{10}M_\odot]$ and $0.42~[\log_{10}M_\odot]$, respectively.
We found empirically that this prior has the same unequal uncertainty levels as constraints from the EM data in Table~\ref{tab:3c66b}. 
Similarly, we fix the luminosity distance based on the redshift of 3C~66B, assuming the $\Lambda$CDM cosmological model based on~\citet{PlanckCollaborationAghanim2020}.
We use an exponential-Normal prior for $\log_{10} f_\text{GW}$, with mean and variance from Table~\ref{tab:3c66b}. 
We perform this calculation with only the Earth term of the CGW.
Finding $\ln \mathcal{B} = - 0.0027(7)$, we claim no evidence for an SMBHB at 3C~66B; nevertheless, we cannot rule out the EM model. 
Finding no evidence, we then calculate upper limits on the parameters of a potential SMBHB in~3C~66B.

Exponentiating our result, the Bayes factor is $\mathcal{B} = 0.9973(6)$, whereas~\citet{ng15_cw_targeted} find a value of $\mathcal{B} = 0.293(3)$. This disagreement is due to noise fluctuations, which are capable of yielding Bayes factors of up to $\sim 1$ according to the null distribution~\citep[see Figure 1, left, in][]{ng15_cw_targeted}.
The use of different pulsars, different instruments, as well as different noise models can influence the fluctuations. 
Additionally, the Bayes factor calculated in this study is based on the Earth term, whereas the Bayes factor in~\cite{ng15_cw_targeted} is calculated based on both the Earth term and the pulsar term. 

\subsection{\label{sec:results:mc} Limits on the Supermassive Black Hole Binary Chirp Mass in 3C~66B}

We perform parameter estimation of the log-10 chirp mass.
As in the previous calculation, we assume the luminosity distance as a fixed parameter based on the known redshift of 3C~66B and Planck 2018 results~\citep{PlanckCollaborationAghanim2020}.
The difference is that now we use a broad uniform prior $\pi(\log_{10} \mathcal{M})=\mathcal{U}(7.5,9.5)~[\log_{10}M_\odot]$. 
This provides a more EM-independent perspective consistent with the decreasing probability of finding an SMBHB toward high $\mathcal{M}>10^9~M_\odot$ in nature at $z=0$~\citep{HopkinsRichards2007}, defined by the SMBHB mass function. 
We also use a more convenient prior on $\pi(\log_{10}(f_\text{GW}))=\mathcal{N}(-7.22,0.01)~[\log_{10}\text{Hz}]$, which is still consistent with the EM model.  

\begin{figure}[h]
     \centering
     \includegraphics[width=\linewidth]{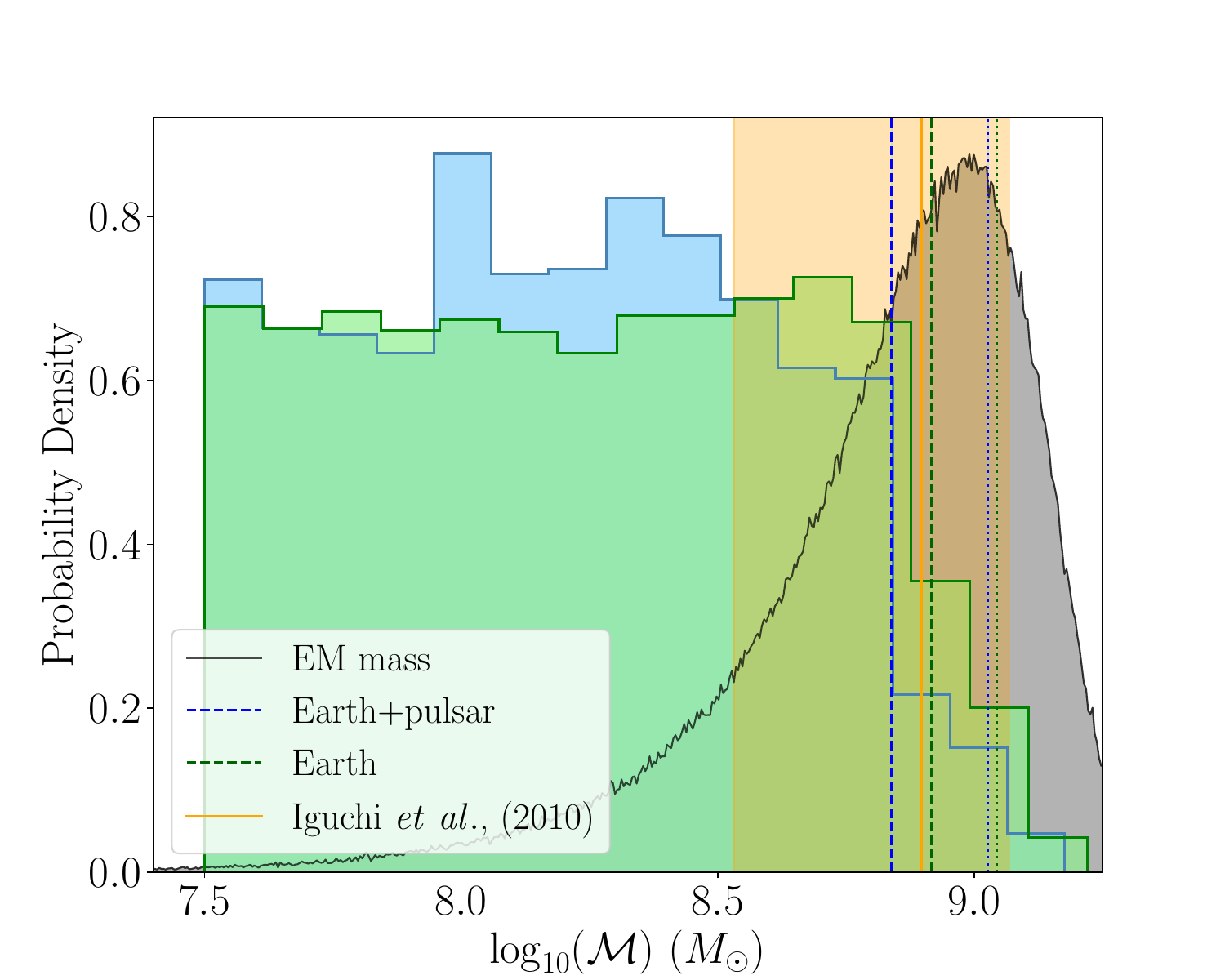}
     \caption{Observational constraints on the chirp mass $\mathcal{M}$ of an SMBHB in 3C~66B.
     The histograms show a posterior for $\mathcal{M}$ based on the analysis of PPTA~DR3. 
     The dashed lines show the respective upper limits at $95\%$ credibility based on the astrophysically motivated prior $\pi(\log_{10}\mathcal{M})=\mathcal{U}(7.5,9.5)~[\log_{10}M_\odot]$. 
     The dotted lines show the limits based on the conservative prior $\pi(\mathcal{M})=\mathcal{U}(10^{7.5},10^{9.5})~[M_\odot]$.
     In green, we show the results of the analysis with only the Earth term of the signal modeled. 
     In blue, we show the results based on modeling both the Earth term and the pulsar term of the CGW. 
     The black solid line with arbitrary density normalization represents the posterior we obtain from the 3 mm data analyzed in~\citet{IguchiOkuda2010}. 
     The vertical orange line and the band correspond to the value and the uncertainty reported in~\citet{IguchiOkuda2010}.
     }
     \label{fig:MassPosteriors}
 \end{figure}

The results are shown in Figure \ref{fig:MassPosteriors}. 
When only the Earth term is modeled in our analysis, we find an upper limit of $\mathcal{M} < 8.25 \times 10^{8}~M_\odot$ at 95\% credibility. 
When including the pulsar term, we find a more constraining limit, $\mathcal{M} < 6.90 \times 10^{8}~M_\odot$. 
For comparison, we show the value from the EM model from~\citet{IguchiOkuda2010}, $(7.9^{+3.8}_{-4.5})\times10^{8}$ $M_{\odot}$. 
Our limits are not below the prediction of the EM model. 
Therefore, we cannot rule it out.
However, the tightest upper limit calculated rules out 58\% of the uncertainty region of the EM-model. 
For consistency with~\cite{ng12_allsky_cw} and \cite{ng15_cw_targeted}, we also obtain a posterior by reweighting our posteriors to a uniform prior in linear space $\pi(\mathcal{M})=\mathcal{U}(10^{7.5},10^{9.5})~[M_\odot]$. 
While not matching astrophysical expectations for the SMBHB mass function, this presents the most conservative limits. 
With this prior, we report an upper limit at $95\%$ credibility to be $\mathcal{M} < 11.06 \times 10^{8} ~M_\odot$ for the Earth term and $\mathcal{M} < 10.63\ \times 10^{8} ~M_\odot$ when the pulsar term is included. 
We find a 16\% improvement in the chirp mass upper limit when including the pulsar term. This difference is on the larger end of the 5-15\% improvement found in previous studies \citep{CharisiTaylor2024, ZhuWen2016}, which compare cases where the signal is detected with a signal-to-noise ratio between 5 and 15. Upper limits use information from the tail of the distribution, whereas detections use the bulk of the distribution, which may explain the larger improvement when including the pulsar term in this work.
Our limits are a factor of 1.55 lower compared to~\citet{ng11_3c66b}, a factor of 1.33 lower compared to~\citet{ng12_allsky_cw}, and similar to~\citet{ng15_cw_targeted}. 

We also report the result of our reanalysis of the 3 mm data from~\cite{IguchiOkuda2010}. 
Following the methodology in Section~\ref{sec:method:3mm}, we obtain a posterior on the log-10 chirp mass of the SMBHB in 3C~66B. 
It is shown as the black shaded area in Figure~\ref{fig:MassPosteriors}. 
The posterior is skewed toward higher masses, consistently with the uncertainties reported in~\cite{IguchiOkuda2010} and presented in Table~\ref{tab:3c66b}.

\subsection{\label{sec:results:h0} Limits on the Gravitational-wave Strain from 3C~66B}

\begin{figure}[h]
     \centering
     \includegraphics[width=\linewidth]{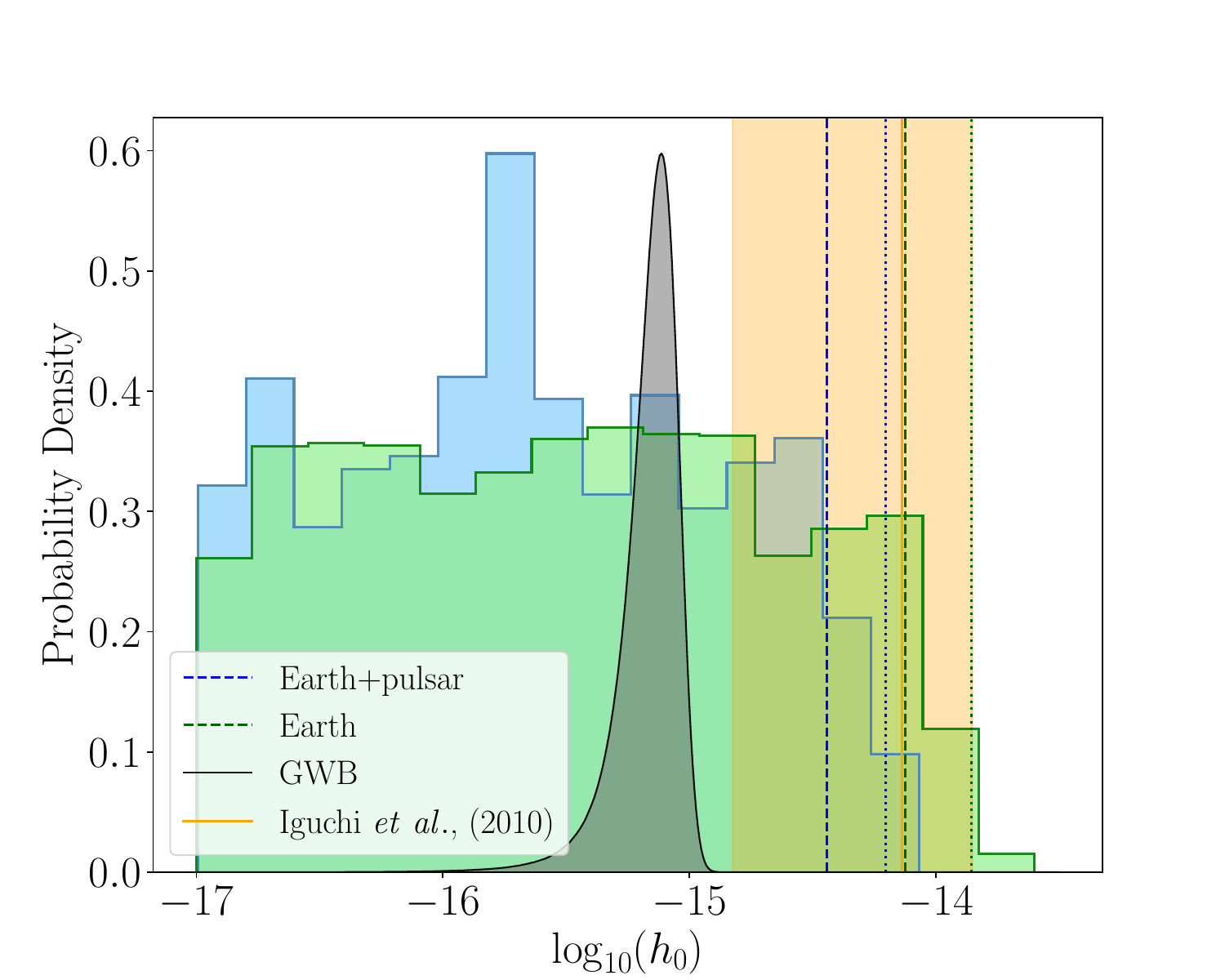}
     \caption{
     Observational constraints on the CGW strain amplitude, $h_0$, of an SMBHB in 3C~66B.
     The histograms show the posterior on $h_0$ based on the analysis of PPTA DR3. 
     The dashed lines show the respective upper limits at $95\%$ credibility based on the astrophysically motivated prior $\pi(\log_{10}h_0)=\mathcal{U}(-17.0,-12.5)$. 
     The dotted lines show the limits based on the conservative prior $\pi(h_0)=\mathcal{U}(10^{-17.0},10^{-12.5})$.
     In green, we show the results of the analysis with only the Earth term of the signal modeled. 
     In blue, we show the results based on modeling both the Earth term and the pulsar term of the CGW.
     The black solid line with arbitrary density normalization represents the theoretically expected GWB contribution at the frequency of 3C~66B.
     The vertical orange line and the band correspond to the value and the uncertainty reported in~\citet{IguchiOkuda2010}.
     }
     \label{fig:hPosteriors}
 \end{figure}

Instead of calculating the strain amplitude $h_0$ of the CGW from 3C~66B based on the luminosity distance using Equation~\ref{eq:strain_amp}, we now treat it as a free parameter.
This way, the distance to 3C~66B is not involved in the calculation.
Similar to the chirp mass prior, we use the prior $\pi(\log_{10}h_0)=\mathcal{U}(-17.0,-12.5)$, which is more consistent with the decreasing probability of an SMBHB with a high $h_0$.
The other priors are as in Section~\ref{sec:results:mc}.
The posterior on $h_0$ is shown in Figure~\ref{fig:hPosteriors}. 
With only the Earth term, we find an upper limit at 95\% credibility to be $\log_{10} h_0 < -14.12$. 
When including the pulsar term, we find a more constraining limit, $\log_{10} h_0 < -14.44$. 
Using the more conservative prior, $\pi(h_0)=\mathcal{U}(10^{-17.0},10^{-12.5})$, we find an upper limit for the Earth term of $\log_{10} h_0 < -13.86$ and an upper limit of $\log_{10} h_0 < -14.20$ when including the pulsar term. As in Section~\ref{sec:results:mc}, the range of $h_0$ of the EM model from~\citet{IguchiOkuda2010} is partially ruled out in all cases.

It is also of interest to compare the strain amplitude of a CGW signal from 3C~66B with the expectation for the GWB at the EM-predicted frequency of the binary. 
\citet{GoncharovSardana2025a} propose a fiducial prior on the characteristic strain amplitude of the background, $h_\text{c}(f) = A(f~\text{yr})^{-2/3}$, characteristic of SMBHBs. Where $A$ is the strain amplitude at the gravitational wave frequency $f$.
The prior is based on black hole masses inferred from kinematic observations and galaxy velocity dispersions. 
Based on this prior, we calculate $h_{0,\text{GWB}}=2h_\text{c}/(2^{2/3}\sqrt{32/5})$, to revert the rms averaging over the SMBHB inclination and polarization assumed for the background (see Equation 8 in~\cite{Burke-SpolaorTaylor2019}). 
The best-fit $h_\text{c}(f=\text{yr}^{-1})$ of the GWB from the data of the European Pulsar Timing Array by~\citet{GoncharovSardana2025a} is within $3\sigma$ interval of the prior, above the prior mean. 
Thus, the prior is consistent with observations. 

The prior from~\citet{GoncharovSardana2025a} marginalized over $\pi(f_\text{CGW})$ is shown as the solid black line in Figure~\ref{fig:hPosteriors}, elucidating a connection between the EM model of the CGW candidate 3C~66B and the expected GWB. 
The comparison suggests the following:
\begin{itemize}
    \item the EM model of 3C~66B corresponds to a loud source in the expected GWB;
    \item a fraction of the parameter space of the EM model for 3C~66B, which we have ruled out, is also not supported by the predicted GWB amplitude; and 
    \item our limits on $f_\text{CGW}$ of 3C~66B are, on average, higher than the expected GWB because we are limited by white noise at the frequency of 3C~66B. 
\end{itemize}
The value of $h_{0,\text{GWB}}$ can also be interpreted as the equivalent expected CGW amplitude of the background if it is dominated by one source (3C~66B) at the 3C~66B CGW frequency. 
This scenario is possible, given the scaling of the number of SMBHB sources contributing to the background with frequency.
This may also lead to a deviation from the power-law model for $h_\text{c}(f)$ at the frequency of 3C~66B, depending on the number density of SMBHBs \citep{Phinney2001, AgazieAnumarlapudi2025}. 
It is also worth noting that~\citet{ZhuCui2019} find that the EM model for 3C~66B is in tension with the observed GWB amplitude based on the expected SMBHB merger rate.

\subsection{\label{sec:results:mass_freq}Upper Limits as a Function of the Continuous Gravitational-Wave Frequency}

\begin{figure}[h]
     \centering
     \includegraphics[width=\linewidth]{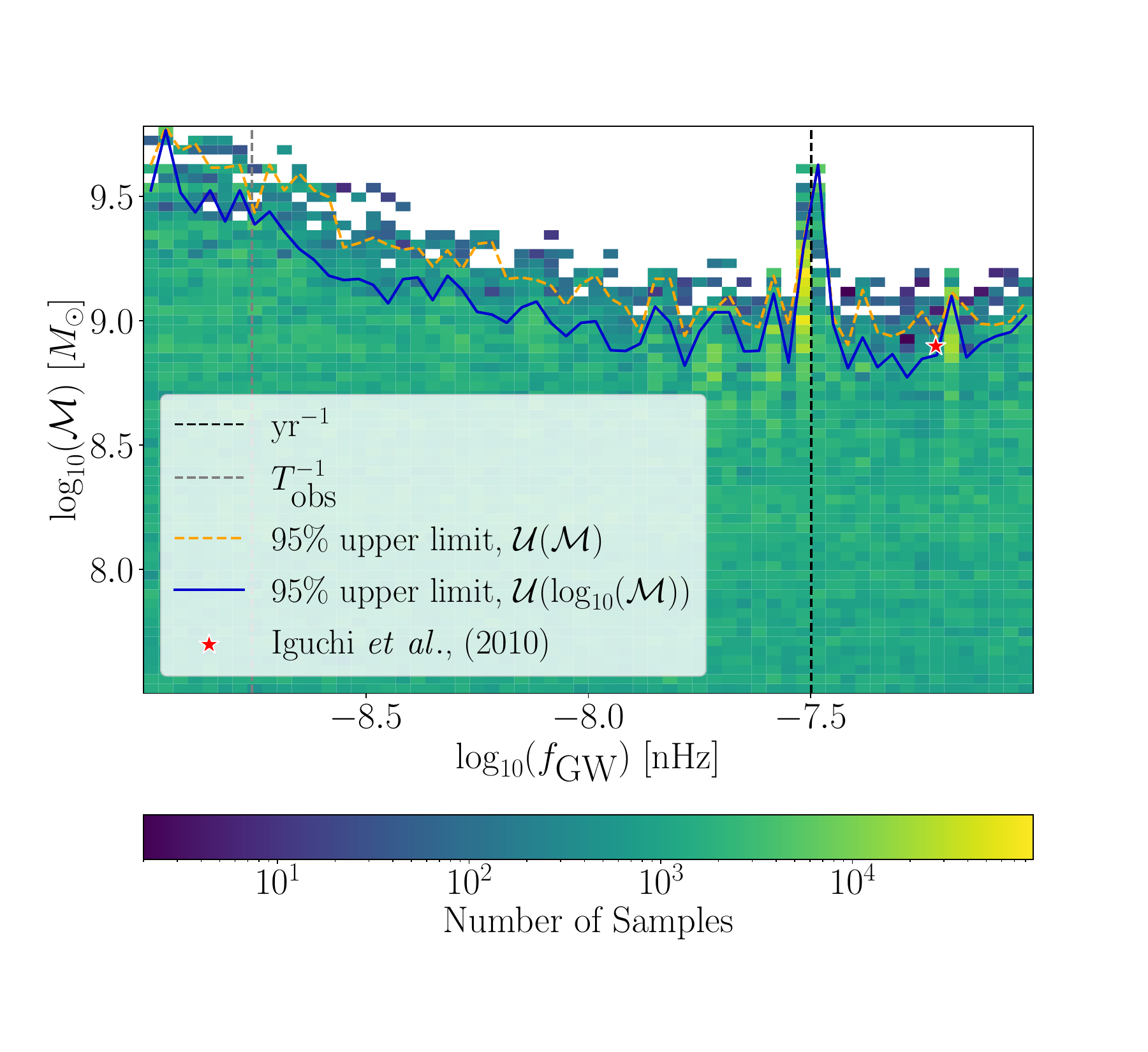}
     \caption{Posterior for the chirp mass $\mathcal{M}$ and frequency $f_\text{GW}$ of 3C~66B.
     The color map shows the number of posterior samples in $\mathcal{M}$-$f_\text{GW}$ space. 
     The best-fit value from~\citet{IguchiOkuda2010} is shown as a red star. 
     The blue solid line shows the respective upper limit at $95\%$ credibility based on the astrophysically motivated prior $\pi(\log_{10}\mathcal{M})=\mathcal{U}(7.5,10.5)~[\log_{10}M_\odot]$. 
     The yellow dashed line shows the limits based on the conservative prior $\pi(\mathcal{M})=\mathcal{U}(10^{7.5},10^{10.5})~[M_\odot]$.
     }
     \label{fig:MassFreq}
 \end{figure}

Following~\citet{ng11_3c66b}, for a broader perspective on how PPTA~DR3 limits on $\mathcal{M}$ depend on $f_\text{GW}$ for any SMBHB at the position of 3C~66B, we show Figure~\ref{fig:MassFreq}.
For this calculation, we use the same prior as in Section~\ref{sec:results:mc}, except the prior on the CGW frequency is now $\pi(\log_{10}f_\text{GW})=\mathcal{U}(-9.0,-7.0)~[\log_{10}\text{Hz}]$. 
We also only model the Earth term of the CGW signal. 
The figure highlights that the PPTA is well suited to target 3C~66B, as the value suggested by the EM model is at the frequency to which our targeted search is most sensitive. 
Consistent with Figure~\ref{fig:MassPosteriors}, when modeling the pulsar term and using astrophysically motivated priors, we are close to ruling out the best-fit value of the EM model. 
We observe the highest density of posterior samples at around $f_\text{GW}=\text{yr}^{-1}$, which corresponds to uncertainties associated with measuring pulsar positions. 
There is also a minor overdensity of posterior samples at $f_\text{GW}=(2~\text{yr})^{-1}$, which corresponds to uncertainties associated with measuring pulsar parallaxes. 
Given that the posterior is not uniform at these frequencies, some of the uncertainties may be systematic.
We find that the chirp mass upper limits are constrained to lower values in the targeted search as compared to the results from Figure 7 in the PPTA DR3 all-sky search \citep{ZhaoChen2025}.

\subsection{\label{sec:results:frequentist} Frequentist Search}

\begin{figure}[h]
     \centering
     \includegraphics[width=\linewidth]{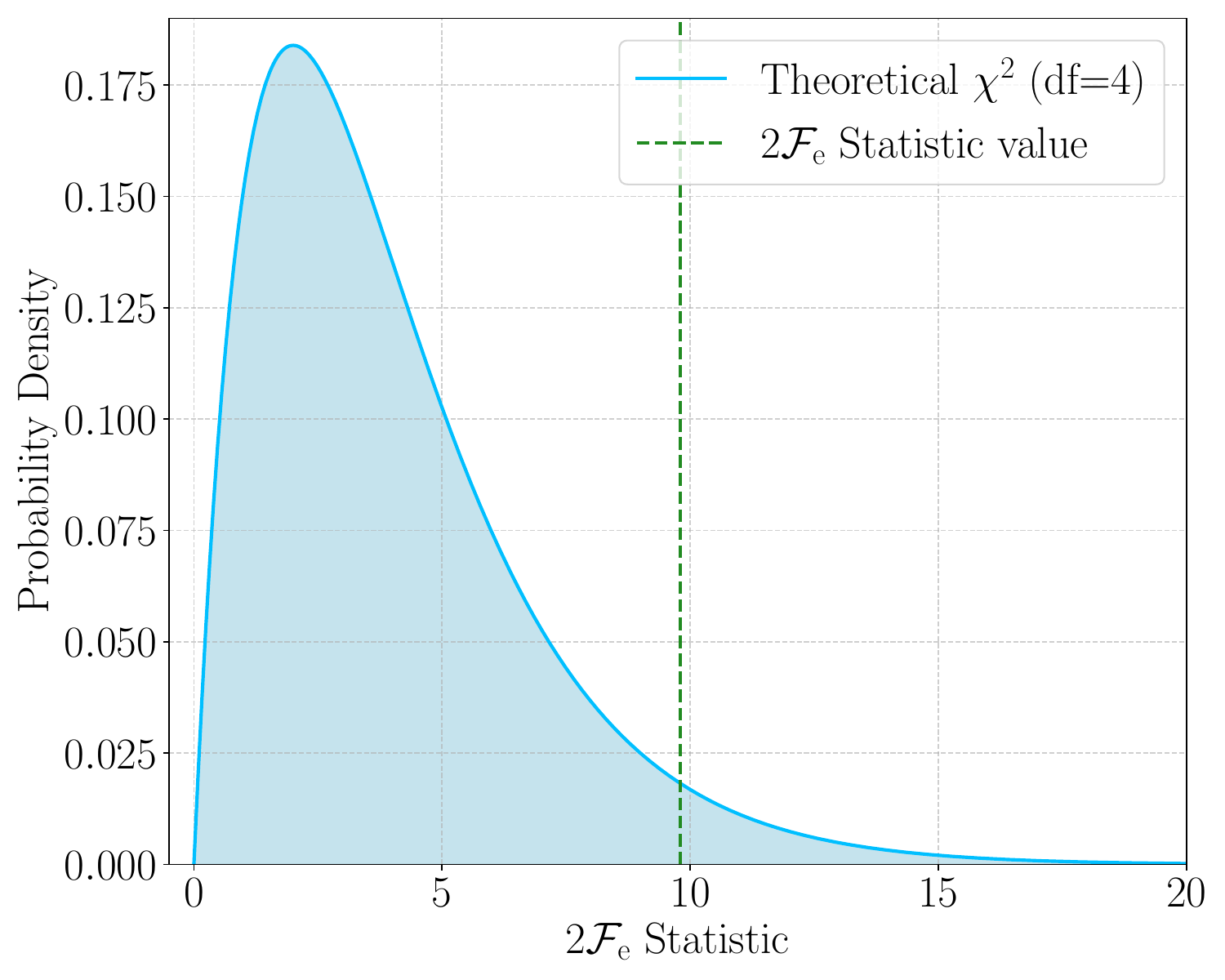}
     \caption{The background distribution of the $\mathcal{F}_\text{e}$ statistic, which is a chi-squared distribution with four degrees of freedom. The green vertical dashed line is the $2\mathcal{F}_\text{e}$ statistic calculated at the frequency and sky position of 3C~66B.}
     \label{fig:FeStat_fulltarget}
 \end{figure}

To confirm the nondetection of a CGW, we perform a frequentist targeted search based on the $\mathcal{F}$ statistic~\citep{EllisSiemens2012}, a frequentist measure of evidence for a CGW in our data. 
In particular, we chose to calculate the $\mathcal{F}_\text{e}$ statistic, which is based on only modeling the Earth term of the CGW signal. 
This statistic is a function of CGW frequency and sky position, which we set to the 3C~66B EM model values.
We find a value of $\mathcal{F}_\text{e}=4.90$, corresponding to the false alarm probability $\text{FAP}=0.50$. 
\citet{EllisSiemens2012} suggest a threshold for FAP to be $10^{-4}$ to claim a detection.
The trials factor for our search is one.
The background distribution of the $\mathcal{F}_\text{e}$ statistic is the chi-squared distribution with four degrees of freedom and is displayed in Figure \ref{fig:FeStat_fulltarget} along with the value of the $2\mathcal{F}_\text{e}$ statistic. 
Based on this null distribution, we find a $p\text{-value} = 0.04$ which corresponds to about $2.1 \sigma$.
Thus, the $\mathcal{F}_\text{e}$ statistic value we find is consistent with noise. 

\section{\label{sec:discussion}Discussion}

\subsection{\label{sec:discussion:multi}Multimessenger Joint Likelihood}

Here, we lay out the groundwork for the joint analysis of CGW and EM data, which may be explored in future work. 
Because EM and CGW data are independent, the total likelihood of these data is a product of the likelihoods: $\mathcal{L}(\bm{\delta t},\bm{S}|\bm{\theta},\bm{\theta}_\text{EM}) = \mathcal{L}(\bm{\delta t}|\bm{\theta}) \mathcal{L}(\bm{S}|\bm{\theta}_\text{EM})$, where $\bm{\theta}=(\bm{\theta}_\text{GW},\mathcal{M})$. 
Effectively, the common parameter in both the CGW model and the EM model from~\citet{IguchiOkuda2010} is the chirp mass $\mathcal{M}$. 
Therefore, the posterior is $\mathcal{P}(\bm{\theta}_\text{GW},\bm{\theta}_\text{EM}, \varphi|\bm{\delta t},\bm{S}) = \mathcal{L}(\bm{\delta t},\bm{S}|\bm{\theta}_\text{GW},\bm{\theta}_\text{EM},\varphi) \pi(\bm{\theta}_\text{GW}) \pi(\bm{\theta}_\text{EM}) \pi(\bm{\varphi})$. 
Here, we also express $\mathcal{L}(\bm{\delta t}|\bm{\theta}_\text{GW},\theta_\text{EM},\varphi)=\mathcal{L}(\bm{\delta t}|\bm{\theta}_\text{GW},\mathcal{M})=\mathcal{G}(\theta_\text{EM},\bm{\varphi})$.

\subsection{\label{sec:discussion:cosmology} Feasibility of Standard Siren Cosmology using Targeted Searches with Pulsar Timing Arrays}

Standard sirens are compact binaries that emit GW and EM radiation simultaneously, allowing for measurement of the Universe's expansion rate~\citep{Schutz1986}. 
The redshift of the host galaxy of a standard siren from EM information enables the determination of its recessional velocity $v$, whereas the proper distance $D$ is measured based on the GW signal. 
The Hubble–Lemaître constant, which quantifies the Universe's expansion at the present time, is 
\begin{equation}
    H_0 = \frac{\nu}{D}.
    \label{eq:H0}
\end{equation}

The primary challenge in the standard siren approach is localizing the EM counterpart of a GW. 
To this date, the first observed binary neutron star inspiral, GW170817~\citep{gw170817}, is the only GW signal for which an EM counterpart is established, for galaxy NGC~4993. 
The sky position uncertainty for GW170817 is $28\ \text{deg}^2$ at 90\% credibility. 
For PTAs, the sky position uncertainty of all-sky CGW searches is expected to be less precise. 
Moreover, the signals are not transient, which makes identification of the host galaxy even more difficult. 
For example, IPTA DR3 simulations (116 pulsars, 20 yr baseline) from \citet{PetrovTaylor2024a} found that for an ideal case where the $\text{SNR} = 15$, the 90\% credible area would range from $29\ \text{deg}^2$ to $241\ \text{deg}^2$. However, for a realistic case, with $\text{SNR} = 8$, the 90\% credible area increases from $287\ \text{deg}^2$ to $530\ \text{deg}^2$. \citet{PetrovTaylor2024a} conclude that without better CGW localization, host galaxy searches are impractical. 
\citet{TruantIzquierdo-Villalba2025} have additionally simulated a 30 yr Square Kilometer Array PTA observation and found that out of the $\sim 35$ CGWs detected, only 35\% have a 90\% credible area $<$~100 $\text{deg}^2$. 
They conclude that observations with an SNR~$>$~25 will have fewer than 100 host galaxy candidates.
Even in simulations which look at a very high SNR, such as those performed by \citet{ZhuWen2016}, with a 15 yr baseline and 30 pulsars, a strong signal with $\text{SNR} = 100$ yields a localization of $\sim40$ $\text{deg}^2$. 
Therefore, standard siren cosmology based on all-sky CGW searches is not feasible with currently operating PTAs. Nevertheless, the use of standard sirens has been explored with simulated future PTA datasets, mostly using dark sirens, and results suggest that precision cosmology may be possible \citep{WangShao2025, YanZhao2020, XiaoShao2025}.  

We propose that by taking advantage of targeted searches for SMBHB candidates, we may be able to revisit the prospect of standard siren cosmology with PTAs.

Current standard siren methods mostly focus on follow-ups from all-sky searches \citep{JinSong2025}, which do not offer \textit{a priori} knowledge of the source position. In contrast, targeted searches based on EM candidates assume the known sky location of the host galaxy. It is important to note, however, that targeted searches do not localize a source within a large uncertainty region, rather they test if the CGW model is consistent with the location of the candidate. This means that source misspecification is possible if there exist multiple hosts that overlap in parameter space. Further studies to determine the extent of the host identification problem and possible solutions are ongoing.

As a demonstration, let us consider a toy example of obtaining constraints on the Hubble–Lemaître constant, provided precise knowledge of the sky position and redshift~\citep{HuntJohnson2021,vandenBoschGebhardt2015} and the (lack of) GW emission from 3C~66B. 
Because $z = 0.02 \ll 0.1$ and because we do not expect stringent constraints on $D$, we assume $D=D_\text{L}$ and use the linear approximation $v=cz$ for simplicity. 
To develop high-precision cosmology with PTAs as part of future work, it is necessary to use full expressions for $D$ and $v$, while also correcting the recession velocities based on peculiar velocities of galaxies~\citep{gw170817_Hubble}.
In any case, the key requirement of a PTA in cosmology, based on targeted CGW searches, is to constrain the luminosity distance of an SMBHB candidate. 

\begin{figure}[h]
     \centering
     \includegraphics[width=\linewidth]{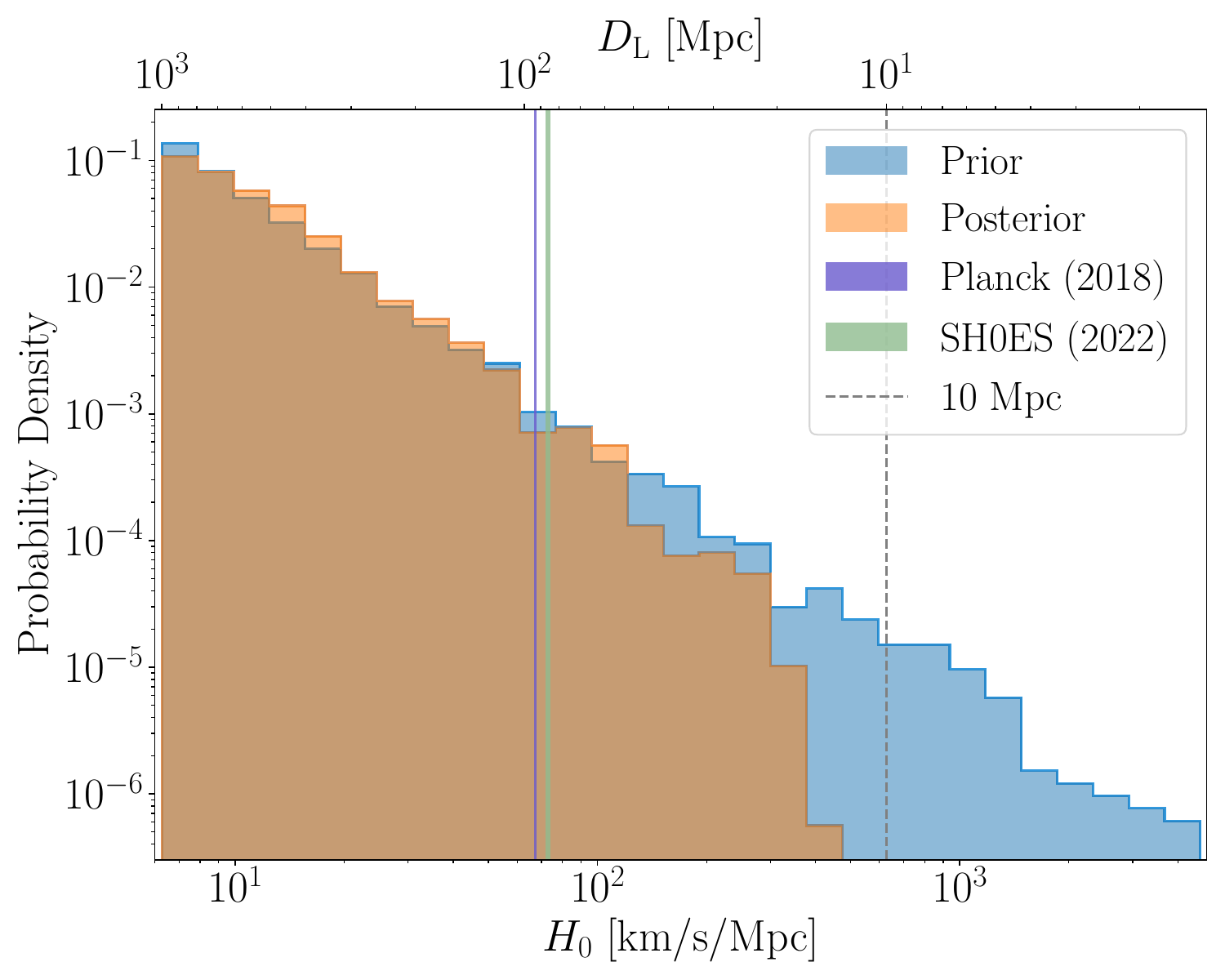}
     \caption{Posterior on the Hubble–Lemaître constant. The prior is shown as the blue histogram and the posterior as the orange histogram. The uncertainty regions for the Hubble–Lemaître constant from \citet{PlanckCollaborationAghanim2020} and the SH0ES collaboration \citep{RiessYuan2022} are plotted as the purple and green regions, respectively. 
     A luminosity distance of 10 Mpc is also plotted for reference as the black vertical dashed line.}
     \label{fig:cosmology_posteriors}
 \end{figure}

Therefore, to obtain a posterior on $H_0$, we place constraints on the luminosity distance $D_\text{L}$ to 3C~66B with the PPTA DR3 data. 
We impose a prior $\pi(D_\text{L})=\mathcal{U}(10^{-2},10^3)$~Mpc. 
We chose a prior on $\log_{10}f_\text{GW}$ as in Section~\ref{sec:results:mc}, and we chose a prior on $\log_{10}\mathcal{M}$ as in Section~\ref{sec:results:model_sel}. 
Based on the posterior on $\pi(D_\text{L})$ and Equation~\ref{eq:H0}, we obtain a posterior on $H_0$ in Figure~\ref{fig:cosmology_posteriors}. 
The values $H_0$ are shown along the bottom horizontal axis. 
The values of $\pi(D_\text{L})$ corresponding to $H_0$ based on Equation~\ref{eq:H0} and the known redshift of 3C~66B are shown along the top horizontal axis. 
Assuming that the EM model is correct, since we do not detect a GW signal, we rule out the nearest luminosity distances. This, in turn, limits the highest values of $H_0$.
We find that the PTA constraints on $H_0$ based on the targeted search generally recover the prior, other than the largest values of $H_0$.
Our choice of $\pi(D_\text{L})$ is presented as a toy model, which is why we do not currently report the $95\%$ credible upper-limit value for $H_0$. 
Because we do not claim any detection of a CGW signal from 3C~66B, we emphasize that these constraints are subject to the validity of the CGW model, where 3C~66B is a true SMBHB. In other words, the EM model for 3C 66B must be correct in order for the constraints to be meaningful. Nevertheless, a strong detection of a CGW from an SMBHB in a targeted search, especially with the resolved pulsar term, may prove to be useful in constraining $H_0$, and investigations into this matter are underway.

\section{\label{sec:conclusion} Conclusion}

Using PPTA DR3, we conduct a targeted multimessenger search for an SMBHB at the center of 3C~66B. 
We search for a CGW signal by performing model selection, where we compare a model with a CGW signal to the null hypothesis. 
With priors carefully matched to the EM model, we find a log Bayes factor of $\ln \mathcal{B} = -0.0027(7)$. 
This suggests we can neither confirm nor rule out the EM model. 
We also find a frequentist statistic, $\mathcal{F}_e = 5.04$, with an associated $\text{FAP}=0.50$, which confirms nondetection of the signal. 
Based on the background estimation, we find that the value of $\mathcal{F}_e = 5.04$ corresponds to $p=0.04$, about $2.1\sigma$. 

Following nondetection, we place upper limits on the mass as $\mathcal{M} < 6.90 \times 10^8~M_{\odot}$  and strain amplitude $\log_{10}(h_0) < -14.44$, at 95\% credibility. 

For both the mass and the strain, we rule out over 50\% of the parameter space of the EM model. 
Using conservative priors, we still rule out a portion of the EM model, finding a $95\%$ upper limit on the chirp mass of $\mathcal{M} < 9.55 \times 10^8~M_{\odot}$ and for the strain amplitude we find a value of $h_0 < -14.20 $. 
Finally, we demonstrate a new methodology for a joint-likelihood approach for the analysis of 3C~66B as well as a proof of principle, showing that it may be possible to use firmly established SMBHB candidates in targeted searches with PTAs as standard sirens, complementary to high-precision cosmology.

\section{Acknowledgments}
Murriyang, CSIRO’s Parkes radio telescope, is part of the Australia Telescope National Facility (https://ror.org/05qajvd42) which is funded by the Australian Government for operation as a National Facility managed by CSIRO. We acknowledge the Wiradjuri people as the Traditional Owners of the Observatory site.
This research has made use of the NASA/IPAC Extragalactic Database, which is funded by the National Aeronautics and Space Administration and operated by the California Institute of Technology.
Parts of this work were supported by the Australian Research Council Centre of Excellence for Gravitational Wave Astronomy (OzGrav) through grant CE230100016

\software{
\textsc{ptmcmcsampler}~\citep{EllisvanHaasteren2019}, 
\textsc{enterprise}~\citep{EllisVallisneri2019}, 
\textsc{enterprise\_warp} at \href{https://github.com/bvgoncharov/enterprise_warp}{github.com/bvgoncharov/enterprise\_warp}, 
\textsc{enterprise\_extensions}~\citep{enterprise_extensions}.
}

\bibliography{mybib,collab}

\appendix

\section{A Continuous Gravitational Wave from a Supermassive Black Hole Binary}
\label{sec:appendix:cw}

Let us start by considering a plane wave traveling from an SMBHB to the SSB to which the pulse arrival times are referenced.
A unit vector that points from the CGW source to the SSB is
\begin{equation}
    \hat{\Omega} = (\phi,\theta) = -\text{sin}\theta \text{cos}\phi\hat{x} - \text{sin}\theta\text{sin}\phi\hat{y} - \text{cos}\theta\hat{z},
\end{equation}
where $(\hat{x},\hat{y},\hat{z})$ is the Cartesian coordinate basis with the north celestial pole in the $\hat{z}$-direction, and the vernal equinox is in the $\hat{x}$-direction. The polar angle $\theta$ and azimuthal angle $\phi$ of the source are related to the equatorial coordinates as $\text{R.A.}=\phi$ and $\text{decl.}=\pi/2-\theta$.

The wave is characterized by the metric strain rank-$2$ tensor in the transverse traceless gauge, indexed with $(a,b)$: 
\begin{equation}
    h_{ab} ( t, \hat{\Omega} ) = e^{+}_{ab} ( \hat{\Omega} )h_+  ( t, \hat{\Omega} ) + e^{\times}_{ab} ( \hat{\Omega} )h_{\times} ( t, \hat{\Omega} ).
\end{equation}
Components $h_{+,\times}$ are the polarization amplitudes, and $e^{+,\times}_{ab}$ are the polarization tensors.
Rotating the Cartesian coordinate system such that it is oriented toward a CGW source, we obtain new basis vectors in $(\hat{x},\hat{y},\hat{z})$ as in \citet{ng12_allsky_cw}:
\begin{equation}
    \begin{aligned} \hat{n}=&(\sin \theta \cos \phi, \sin \theta \sin \phi, \cos \theta) = -\hat{\Omega}, \\ \hat{p}=&(\cos \psi \cos \theta \cos \phi-\sin \psi \sin \phi,\\ &\cos \psi \cos \theta \sin \phi+\sin \psi \cos \phi,-\cos \psi \sin \theta), \\ \hat{q}=&(\sin \psi \cos \theta \cos \phi+\cos \psi \sin \phi,\\ &\sin \psi \cos \theta \sin \phi-\cos \psi \cos \phi,-\sin \psi \sin \theta).
    \end{aligned}
\end{equation}
The polarization angle $\psi$ is the angle between the natural polarization basis (at source) and the reference polarization basis (at SSB).
With this, polarization tensors are expressed as 
\begin{equation}
    \begin{aligned}
        e^{+}_{ab} = \hat{p}_a\hat{p}_b - \hat{q}_a\hat{q}_b, \\
        e^{\times}_{ab} = \hat{p}_a\hat{q}_b + \hat{q}_a\hat{p}_b.
    \end{aligned}
\end{equation}

The response of a PTA to the plane wave is described by the
antenna pattern function~\citep{RomanoCornish2017}
\begin{equation}
    F^{A}(\hat{\Omega}) \equiv \frac{1}{2} \frac{\hat{u}^{a} \hat{u}^{b}}{1+\hat{\Omega} \cdot \hat{u}} e_{a b}^{A}(\hat{\Omega}),
\end{equation}
where $\hat{u}$ is the unit vector pointing from the SSB to the pulsar.
This function describes the pulsar's geometrical sensitivity to the CGW source \citep{TaylorHuerta2016}. Pulsar timing delays $\delta t$ are then expressed as
\begin{equation}
    \delta t(t, \hat{\Omega})=F^{+}(\hat{\Omega}) \delta t'_{+}(t)+F^{\times}(\hat{\Omega}) \delta t'_{\times}(t),
\end{equation}
where we implied that $\delta t_{+,\times}$ is the time integral of $h_{+,\times}$, and $\delta t'_{+,\times}$ is the difference between the delay experienced at the SSB (Earth) and the signal induced at the pulsar. This difference is 
\begin{equation}
    \delta t'_{+, \times}(t)=\delta t_{+, \times}(t_\text{p})-\delta t_{+, \times}(t),
\end{equation}
where $t$ is the time when the CGW passes the SSB and $t_\text{p}$ is the time when the CGW passes the pulsar. 
One can then geometrically relate $t$ and $t_\text{p}$, as
\begin{equation}
    t_{p}=t-\frac{L}{c}(1+\hat{\Omega} \cdot \hat{u}),
\end{equation}
where $L$ is the distance to the pulsar. 

For an SMBHB, at zeroth post-Newtonian order, and choosing a coordinate system where $\psi = 0$, we arrive at our model of the CGW projected onto the PTA detector: 
\begin{equation}
\begin{aligned} \delta t'_{+}(t)=& \frac{(G \mathcal{M})^{5 / 3}}{D_\text{L} c^4 \omega(t)^{1 / 3}}[-\sin 2 \Phi(t)(1+\cos ^{2} \iota)], \\ 
\delta t'_{\times}(t)=& \frac{(G \mathcal{M})^{5 / 3}}{D_\text{L} c^4 \omega(t)^{1 / 3}}[2 \cos 2 \Phi(t) \cos \iota ], \end{aligned}
\end{equation}
where $\iota$ is the SMBHB inclination angle, $D_\text{L}$ is the luminosity distance, $\omega(t)=\pi f_\text{GW}$ is the angular orbital frequency\footnote{The SMBHB angular orbital frequency is half the CGW angular frequency $2 \pi f_\text{GW}$.}, $\Phi  ( t  )$ is the phase, and
\begin{equation}
    \mathcal{M} \equiv  \frac{(m_{1} m_{2})^{3 / 5}}{(m_{1}+m_{2})^{1 / 5}}
\end{equation}
is the chirp mass, where $m_{1,2}$ are the SMBHB component masses.
The last observation in the PPTA dataset (2022 March 8, or MJD 59646) is used as the reference time. 
The orbital phase and frequency of the SMBHB are then defined as
\begin{equation}
    \begin{array}{l}\Phi(t)=\Phi_{0}+\frac{1}{32} (\frac{G\mathcal{M}}{c^3})^{-5 / 3}[\omega_{0}^{-5 / 3}-\omega(t)^{-5 / 3}], \\ \omega(t)=\omega_{0}(1-\frac{256}{5} (\frac{G\mathcal{M}}{c^3})^{5 / 3} \omega_{0}^{8 / 3} t)^{-3 / 8},\end{array}
\end{equation}
with $\Phi_0$ being the initial orbital phase and $\omega_0$ the initial orbital frequency; one may notice that this equation is equivalent to Equation~\ref{eq:freqev}. 

The chirp mass and frequency are redshifted for CGW sources at cosmological distances, such that the observed (``$o$'') values are related to intrinsic source (``$s$'') values as 
\begin{equation}
    \begin{aligned} \mathcal{M}_\text{o} &=\mathcal{M}_\text{s}(1+z), \\ 
    f_o &=\frac{f_s}{1+z}. \end{aligned}
\end{equation}
For 3C~66B, $z=0.02 \ll 1$, so given our measurement uncertainty we assume $\mathcal{M} = \mathcal{M}_o = \mathcal{M}_s$ and $f=f_o=f_s$.

\section{Analysis of the 3 mm Data}
\label{sec:appendix:em}

Although key details of the analysis of the 3 mm data are provided in~\citet{IguchiOkuda2010}, here we provide explicit equations that we used in our reanalysis. 
Throughout, we switch between the velocity $v$ and the boost parameter $\beta=v/c$, which may also have subscripts ``bl,0'' and ``app'' for the absolute binary component velocity around the center of gravity and the apparent relativistic jet velocity, respectively. 
First, we point out that Equations 1 and 2 from~\citet{IguchiOkuda2010} can be expressed directly in terms of $(\beta,\iota)$, as
\begin{equation}
    \begin{cases}
    \beta = \sqrt{(1-r)^2 + (\beta_\text{app}r)^2},\\
    \iota = \text{atan2}(\beta_\text{app}r, 1-r),
    \end{cases}
\end{equation}
where $r$ is the ratio
\begin{equation}
    r = \frac{P_\text{obs}}{(1+z)P_\text{k}}.
\end{equation}
Next, $(\beta,\iota)$ are required in the calculation of the absolute value of the rotational velocity of one of the binary components, which has a jet, around the center of mass. 
This velocity causes Doppler boosting of the flux.
We solve Equation 4 from~\citet{IguchiOkuda2010} for this velocity, obtaining 
\begin{equation}
    \beta_\text{bl,0} = \frac{1 - \beta \cos\iota}{\sin\iota} \frac{S_\text{r}^{\frac{1}{3+\alpha}}-1}{S_\text{r}^{\frac{1}{3+\alpha}}+1}.
\end{equation}
Considering Keplerian motion in a circular orbit, we use the following expressions of the orbital radius $R$ and the orbital separation $a$,
\begin{equation}
    \begin{cases}
    a = \sqrt[3]{\frac{G M_\text{tot} P^2_\text{k}}{(2 \pi)^2}},\\
    R = \frac{v_\text{bl,0} P_\text{k}}{2 \pi}.
    \end{cases}
\end{equation}
Introducing $x \equiv R/a$, one finds that $x=m/M_\text{tot}$ with $m$ the mass of the lighter black hole for the case where a heavier black hole emits a jet (smaller radius). 
Similarly, one finds $x=M/M_\text{tot}$ with $M$ the mass of the heavier black hole for the case where the lighter black hole emits a jet (larger radius). The total mass, $M_{\text{tot}}$, is obtained by using the relationship between the central stellar velocity dispersion and black hole masses \citep{FerrareseMerritt2000, MerrittFerrarese2001}. Using this relationship \citet{Noel-StorrBaum2007} obtain a total mass for 3C 66B of $1.9^{+1.0}_{-0.8} \times 10^{9}\ M_{\odot}$.
Apparently, for the calculation of the chirp mass, these two cases reduce to the same answer due to a symmetry in $x(1-x)$ in the final expression
\begin{equation}
    \begin{cases}
    \mathcal{M} = M_\text{tot}[x(1-x)]^{3/5},\\
    x \equiv \frac{R}{a} = v_\text{bl,0} \sqrt[3]{\frac{P_\text{k}}{2 \pi G M_\text{tot}}}.
    \end{cases}
\end{equation}
The value of $x(x-1)$ is also known as the symmetric mass ratio.

\section{\label{sec:appendix:pdist}Pulsar Distances}

Pulsar distances are required in the calculation of the expected signal when modeling the pulsar term of the CGW signal. 
Pulsar distances are measured using very long baseline interferometry (VLBI) astrometry, optical astrometry, timing, and the \citet{Shklovskii1970} effect. 
A list of the priors used for each pulsar, as well as the technique used to find this distance, are listed in Table~\ref{tab:pulsar_distances}. 
We impose a Gaussian prior on pulsar distance in accordance with the estimates found in the literature. 
When no such estimates are available for a given pulsar, we impose a sufficiently conservative prior $\mathcal{N}(1000,200)~\text{[pc]}$. 
Distance estimates based on pulsar dispersion measures are not used due to the large uncertainties in the electron density model~\citep{YaoManchester2017}. 

\begin{table*}[]
\renewcommand{\arraystretch}{1.3}
\centering
\caption{Table Describing the Pulsar Distances, the Methods Used to Obtain Them, and the Source from Which They Were Obtained.}
\begin{tabular}{cccc}
\hline
PSR & Distance [pc]                     & Measurement             & Reference \\ \hline
 & (pc)                     &              & \\ \hline \hline
J0030+0451  & $329^{+6}_{-5}$               & VLBI astrometry                   & \citet{DingDeller2023}       \\
J0125-2327  & $1000 \pm 200$                 & ...                                 & \citet{EllisVanHaasteren2017}       \\
J0437-4715  & $156.96\pm0.11$               & Shklovskii effect                 & \citet{Shklovskii1970, ReardonBailes2024}       \\
J0613-0200  & $125\pm54.69$                 & Timing                            & \citet{VerbiestBailes2009}       \\
J0614-3329  & $1000 \pm 200$                 & ...                                 & \citet{EllisVanHaasteren2017}       \\
J0711-6830  & $1000 \pm 200$                 & ...                                 & \citet{EllisVanHaasteren2017}       \\
J0900-3144  & $1000 \pm 200$                 & ...                                 & \citet{EllisVanHaasteren2017}       \\
J1017-7156  & $256.41\pm78.90$              & Timing                            & \citet{NgBailes2014}       \\
J1022+1001  & $720.98^{+21.31}_{-14.55}$    & VLBI astrometry                   & \citet{DellerGoss2019}       \\
J1024-0719  & $1072^{+67}_{-49}$            & Optical astrometry                & \citet{MoranMingarelli2023}       \\
J1045-4509  & $303.03\pm 174.47$            & Timing                            & \citet{VerbiestBailes2009}       \\
J1125-6014  & $1000 \pm 200$                 & ...                                 & \citet{EllisVanHaasteren2017}       \\
J1446-4701  & $1000 \pm 200$                 & ...                                 & \citet{EllisVanHaasteren2017}       \\
J1545-4550  & $1000 \pm 200$                 & ...                                 & \citet{EllisVanHaasteren2017}       \\
J1600-3053  & $5000 \pm 3750$               & Timing                            & \citet{VerbiestBailes2009}       \\
J1603-7202  & $1000 \pm 200$                 & ...                                 & \citet{EllisVanHaasteren2017}       \\
J1643-1224  & $4763.36^{+99.06}_{-104.89}$  & VLBI astrometry                   & \citet{DingDeller2023}       \\
J1713+0747  & $1052.63^{+66.48}_{-55.40}$   & VLBI astrometry                   & \citet{ChatterjeeBrisken2009}       \\
J1730-2304  & $510^{+30}_{-30}$             & VLBI astrometry                   & \citet{DingDeller2023}       \\
J1744-1134  & $416.67 \pm 17.36$            & Timing                            & \citet{VerbiestBailes2009}       \\
J1824-2452A & $10000 \pm 50000$             & VLBI astrometry                   & \citet{DingDeller2023}       \\
J1832-0836  & $1000 \pm 200$                 & ...                                 & \citet{EllisVanHaasteren2017}       \\
J1857+0943  & $909.09 \pm 165.29$           & VLBI astrometry                   & \citet{VerbiestBailes2009}       \\
J1902-5105  & $1000 \pm 200$                 & ...                                 & \citet{EllisVanHaasteren2017}       \\
J1909-3744  & $1265.82 \pm 32.05$           & Timing                            & \citet{VerbiestBailes2009}       \\
J1933-6211  & $1000 \pm 200$                 & ...                                 & \citet{EllisVanHaasteren2017}       \\
J1939+2134  & $2604.17^{+325.52}_{-311.96}$ & VLBI astrometry                   & \citet{DingDeller2023}       \\
J2124-3358  & $322.58 \pm 57.23$            & Timing                            & \citet{VerbiestBailes2009}       \\
J2129-5721  & $2358.49 \pm 489.50$          & Optical astrometry           & \citet{JenningsKaplan2018}       \\
J2145-0750  & $623.83^{+24.51}_{-3.50}$     & VLBI astrometry                   & \citet{DellerGoss2019}       \\
J2241-5236  & $1000 \pm 200$                 & ...                                 & \citet{EllisVanHaasteren2017}   \\ \hline   
\end{tabular}
\vspace{2mm}
\begin{minipage}{\textwidth}
\centering
\footnotesize
Note. If no distance was available, the method is listed as ``...".
\end{minipage}
\label{tab:pulsar_distances}
\end{table*}

\end{document}